\documentstyle[12pt,epsfig]{article}
\renewcommand{\title}[1]{{\Large\bf\mbox{}\\\medskip#1\bigskip\medskip\\}}
\renewcommand{\author}[1]{{\large #1\smallskip\\}}
\newcommand{\address}[1]{{\em #1\medskip\\}}
 
\def\be{\begin{equation}}
\def\ee{\end{equation}}
\def\bea{\begin{eqnarray}}
\def\eea{\end{eqnarray}}
\def\ba{\begin{array}}
\def\ea{\end{array}}

\def\0{$\Gamma_0$}
\def\o{\omega}

\begin{document}
\begin{center}

\title{Zeroes of the Jones polynomial}
\author{F. Y. Wu and J. Wang}
\address{Department of Physics\\
Northeastern University, Boston, Massachusetts 02115, U. S. A.}


\begin{abstract}
We study the distribution of 
zeroes of the Jones polynomial $V_K(t)$ for a knot $K$.
We have computed  numerically the roots of the Jones polynomial for
all prime knots with $N\leq 10$ crossings, and found 
the zeroes scattered about the unit circle
$|t|=1$ with the average distance to the  circle approaching
 a nonzero value as $N$ increases. 
  For  torus knots of the type $(m,n)$ we show that
 all zeroes lie on the unit circle with
a uniform density in the limit of either $m$ or
 $n\to
\infty$, a fact confirmed by our numerical findings. We have
also elucidated the relation connecting the Jones polynomial with the
Potts model, and used this relation to derive
the Jones polynomial for
  a repeating chain knot with $3n$ crossings for general $n$.  
It is found that 
 zeroes of its Jones polynomial  lie
on three closed curves centered about the  points $1, i$ and $-i$.
In addition, there are  two isolated zeroes   located one each
near the points  $t_\pm = e^{\pm 2\pi i/3}$
at a distance of the order of
$3^{-(n+2)/2}$.    
 Closed-form
expressions are deduced for the  closed curves in the limit of $n\to \infty$.

\end{abstract}
\end{center}


\vskip 1cm

\section{Introduction}
A powerful tool often used in the study of lattice models in statistical
mechanics is the consideration of  zeroes of the partition function.
Since partition functions of lattice models are usually  of the form of
 polynomials, and  polynomials are completely determined by their roots,
 a study of its zero distribution   often leads to insights
to physical properties  otherwise difficult to see.
   For the  ferromagnetic Ising model, for example, one has  the remarkable Yang-Lee circle theorem
\cite{yanglee, leeyang} in the complex magnetic plane
 which  rules out the existence of phase transitions in a nonzero field.
   Similarly, 
zeroes of the Ising partition function in the complex temperature 
plane leads to Fisher circles \cite{fisher},
a consideration  yielding
 information on the nature of the  transition.

\medskip
As the occurrence of polynomials is also  common in mathematics,
 it is  natural to inquire whether useful information
can be gained by studying zeroes of these mathematical entities.
An example of such a consideration is
the   restricted $d$-dimensional partition
of an integer \cite{andrews}.
While  the partition generating function,
which is always of the form of a polynomial, is known 
for $d=2,3$   \cite{macmahon},
the $d\geq 4$ analysis
has remained a well-known intractable problem in number theory 
for over one century \cite{andrews}. 
By studying the  zeroes
of the generating function for $d=4$,  however, some regularity in the distribution 
has been found \cite{huangwu}, thus giving rise to the hope
that there may indeed be some truth lurking behind
the scene \cite{andrews1}.

\medskip
 In this {\it Letter} we consider zeroes of   the 
Jones polynomial, the algebraic invariant for knots and links \cite{jones}.
 It is  well-known that the Jones polynomial is
related to the Potts model \cite{jones} and  that it 
 is given by
the Potts partition function with a special parametrization  \cite{jones2,wureview}.
 Now  the Potts model
 has been analyzed extensively by evaluating its partition function zeroes
(see, for example, \cite{maillard,wupotts,shrock} 
for zeroes of the Potts partition function in the complex temperature plane 
  and  \cite{baxter, shrocktsai} for zeroes of the related chromatic polynomial).
 It is therefore  of  interest to extend the  zero consideration to the Jones polynomial,
  an approach which does not appear to have  been
taken previously.

\medskip
The organization of this paper is as follows.
  In Sec. 2 we outline the connection
between the Jones polynomial and the Potts model, and write down an explicit relation
connecting the Jones polynomial to a Potts partition function, which does not appear to have
been given previously.
In Sec. 3 we present numerical results 
  for the zeroes of the Jones polynomial for
prime knots with 10 or less crossings.  In Sec. 4 we consider the
$(m,n)$ torus knot, and show in particular that all zeroes lie on the unit
circle with a uniform density 
in the limit of either $m$ or $n\to \infty$.  Using
the Potts model formulation given in Sec. 2,
we derive in Sec. 5
the Jones polynomial for a repeating chain knot having $3n$ crossings
for general $n$, and 
show that, in the limit of $n\to \infty$,  its zeroes are located at
two isolated points and on three closed curves along which the Jones polynomial is
nonanalytic.  The exact location of the closed curves are determined.

\section{The Jones polynomial and the Potts model}
In this section we describe 
the connection of the Jones polynomial with the
Potts model, and write down the explicit relation connecting the two, a result we use in Sec. 
5.
 
\medskip
 The  partition function of a $q$-state Potts model \cite{potts} 
with spins residing at sites of  a graph $G$ and
 interactions
$K_{ij}$ along the edge  connecting sites $i$ and $j$ is given by the summation
\be
Z(q;\{e^{K_{ij}}\}) = \sum_{\sigma_i=1}^q \prod_{<i,j>}B(\sigma_i, \sigma_j). \label{part}
\ee
Here, $\sigma_i=1,2,\cdots,q$ denotes the spin state of the site $i$,
the product is taken over all edges of $G$,
 and the Boltzmann factor is
\be
B(\sigma_i, \sigma_j) = 1+(e^{K_{ij}}-1)\delta_{\rm Kr}(\sigma_i, \sigma_j). 
\label{boltzmann}
\ee
 Clearly, the partition function is a multinomial in $q$ and $e^{K_{ij}}$.
 
\medskip
In physics one is mostly interested in the evaluation of 
the partition function (\ref{part}) for large
lattices $G$, and focuses on the nature  of its 
singularity  associated with phase transitions.
  But  the Potts partition function  
plays   equally important roles in 
mathematics  where the emphasis is
on  finite planar graphs $G$. It is well-known that the Potts partition function is
completely equivalent to the 
Tutte polynomial \cite{tutte} arising in graph theory.  By
taking $e^{K_{ij}}=0$,  for example, the partition function $Z(q, \{ 0 \})$ gives
  the chromatic polynomial for the number of $q$-colorings
of $G$ \cite{birkhoff}.  The Potts model is also
related to the Jones polynomial in the knot theory via the thread of the
Temperley-Lieb algebra
  \cite{jones2,tl}.  The exact relation  connecting the Jones polynomial
to a Potts partition function is given below.
   
\medskip
Let $V_K(t)$ be the Jones polynomial of 
 an oriented knot with a  planar projection $K$.
There are  
two kinds of line crossings, $+$ and  $-$, in $K$, and let 
  $n_+$ and $n_-$ be the respective numbers of  crossings.
Shade alternate faces of $K$ (there are two ways to do this), and the
  shading of faces  yields two
kinds of line crossings determined solely by the relative positioning of the
shaded faces  with respect to the topology of the line crossing, without regarding to the
line orientation. Specifically,
the  crossing is $+$ ($-$)
   if  the shaded area appears on the left (right)
when traversing on an ``overpass" at the crossing \cite{jones1,wureview}.
 
\medskip
Place a Potts spin in  each of the
shaded faces and let the spins  interact  along edges drawn
 over the line crossings.
The interaction is $K_+$ ($K_-$) if the crossing is $+$ ($-$) as determined
by the face shading.
 Let the number of  $K_\pm$ interactions be $m_\pm$,
 and let $Z(q;e^{K_+},e^{K_-})$ be the Potts partition function.
 Then, we have the following exact equivalence\footnote{The expression (\ref{V}) is the same as
Eq. (7.17) of \cite{wureview} after 
correcting a misprint of $t^{3/4} \to t^{-3/4}$.}
\be
V_K(t) = q^{-(M+1)/2} (-t)^{-N} t^{(3n_+ + m_+)/2}Z(q;e^{K_+}, e^{K_-})
\label{V}
\ee
where   $M$ is the total number 
of  Potts spins, $N=n_++n_-=m_++m_-$ is the number of crossings in $K$, and
 \be
q = t + 2 + t^{-1}, \hskip 1cm e^{K_+} = -t^{-1}, \hskip 1cm e^{K_-} = -t. 
\ee

\medskip
If one chooses to shade the other set of faces of $G$,
the topology consideration then
 interchanges  $K_+ \leftrightarrow K_-$,
and hence $m_+ \leftrightarrow m_-$, resulting in a Potts model on a dual graph.
One then establishes, by the use of
a duality relation of the Potts partition function \cite{wuwang},
that this leads to the same Jones polynomial (\ref{V}) with an extra sign $(-1)^N$.

 \section{Prime knots}
We first study the  zeroes for prime knots with specific numbers of
crossings.
 
\medskip
It has been shown by Jones \cite{jones1} that
the Jones polynomial $V_K(t)$ for a knot $K$ assumes the general form 
\be
V_K(t) = 1- (1-t)(1-t^3)W_K(t) \label{jonesk}
\ee
where $W_K(t)$ is some Laurent polynomial.  Using the explicit expression 
of $W_K(t)$ 
tabulated by Jones, 
 we have computed the zeroes of $V_K(t)$ for all prime knots   with $N\leq 10$
crossings.   The results for $N=7,8,9,10$ are shown  in Fig. 1. 
(Results for $N\leq 6$ are not included since there are only 7 knots 
in this category whose
  statistical importance is insignificant.)
  It is found  that the roots are scattered about the unit circle $|t|=1$,
with a  nonzero  fraction $F$ of the roots residing on the real axis.
 As a quantitative measurement of the distribution, we have computed
the average distances
 of the zeroes from the unit circle for each $N$,
 and obtained the results shown  in 
Table 1.  Here, $d_1$ is the average distance from the unit circle for
all roots, real and complex, and $d_2$ is the average distance for complex roots only.
In both cases, the average distance appears to settle at
 $d_1 \sim 0.22, d_2\sim 0.17$ as $N$ increases.

\begin{center}
\begin{tabular}{cccccc} \\
\hline\hline
$N$ & \# of knots &  \# of roots & F& $d_1$ & $d_2$  \\
\hline
7 & 7 &54 &0.22&0.2610 & 0.1708   \\
8  &21 &160 & 0.11 &0.2061 & 0.1684  \\
9 &49 &  440& 0.15 &0.2277 & 0.1704 \\ 
10 &166 & 1,585& 0.14  &0.2179  & 0.1729  \\
 \hline\hline
\end{tabular}

{Table 1. Average distance of zeroes from the unit circle for prime knots.}
\end{center}

\medskip
The two points $t_\pm = e^{\pm 2\pi i/3}$ on the unit circle
appear to be special since there is an
apparent clustering of zeroes in the neighborhoods.
 This fact can be understood as follows.  
>From (\ref{jonesk}) we have 
 $V_K(t_\pm) =1$   for all $K$
and $N$ finite.
 Consider next whether one has a solution at $V_K(t_\pm + \Delta)=0$
for some small $|\Delta|$.  To the leading order of $|\Delta|$, (\ref{jonesk})
yields
\be
|\Delta| \sim {1\over {3\sqrt 3 |W_K(t_\pm)|}},
\ee
if $W_K$ is slow-varying in $t$ near $t_\pm$. 
 In practice, one finds $|W_K(t_\pm)|$ always of the order of 1 or larger (except that 
it vanishes identically for the   knots $7_2, 8_1, 9_5, 10_3, 10_{20}, 10_{34}$,
and $ 10_{135}$).
  It follows that we have $|\Delta|\sim 0.1$  consistent
with the assumption of a small $|\Delta|$.  Indeed, the 
existence of this solution for each knot
is verified numerically.
 As each Jones polynomial contributes one  root near $t_+$ (and $t_-$),
 the roots then appear to
  cluster near $t_\pm$ as shown.
 
\begin{figure}[htbp]
\center{
\epsfig{figure=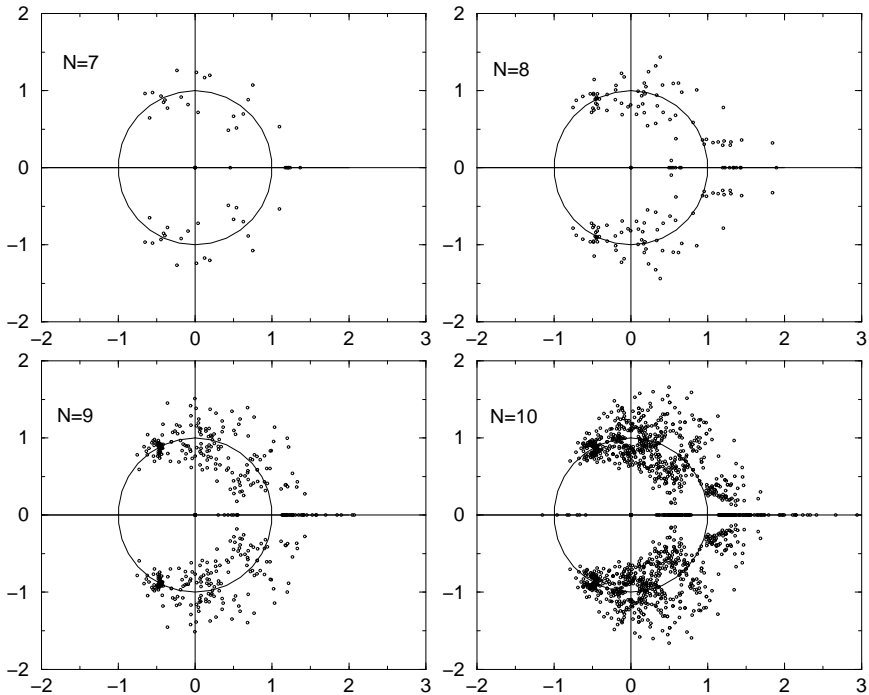,height=4.5in,angle=-90}}
\vskip -.2in
\caption{Distribution of zeroes for prime knots with $N=7,8,9,10$ crossings.}
\label{fig1}
\end{figure}

\section{Torus knots}
A torus knot of type $(m,n)$ is 
the closure of an $m$-line braid with $n$-step permutations,
when $m$ and $n$ are relative primes
\cite{jones1}.  For the $(m,n)$ torus knot the Jones polynomial is \cite{jones1}
  \be
V_K(t) = 
 { {t^{(m-1)(n-1)/2}} \over { (1-t^2) }     } 
\Big(1-t^{m+1} - t^{n+1} + t^{m+n}\Big),  \label{torus}
\ee
which is symmetric in $m$ and $n$.
Setting the numerator equal to zero and rearranging, we obtain
 \be
t^n =  {1\over t} \Bigg({{1-t^{m+1}}\over {1-t^m}} \Bigg),\label{torus1}
\ee
which yields, after  taking the $n$-th root, 
\be
t_\alpha =  \Bigg|{1\over t} \Bigg({{1-t^{m+1}}\over {1-t^m}} \Bigg)\Bigg|^{1/n}\o^\alpha ,
\hskip 1cm \alpha = 1,2,..., n,
\ee
where $\o = e^{2\pi i/n}$. 
Taking the limit of $n\to\infty$, this leads to
$t_\alpha \to   \o^\alpha$ for all $m$,
indicating that  all  roots are distributed uniformly on the
unit circle $|t|=1$, a fact valid for any $m$.
Numerical solutions of (\ref{torus1}) for
finite $(m,n)$ which borne out this fact are shown in Fig. 2. 

\begin{figure}[htbp]
\center{
\epsfig{figure=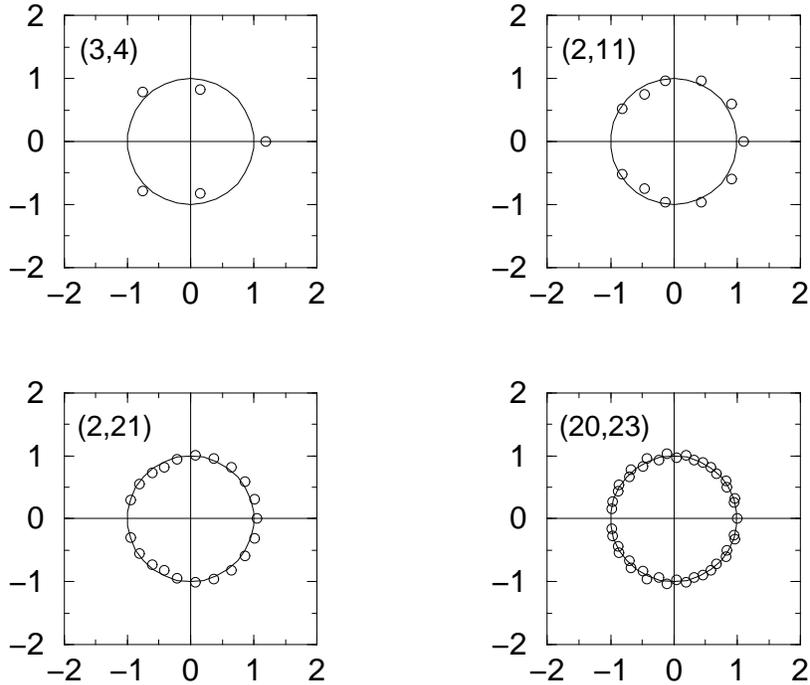,width=4.2in}}
\vskip -3.4in
\caption{Distribution of zeroes for the torus knot.}
\label{fig2}
\end{figure}

\section{A repeating chain knot}
In this section we consider
the Jones polynomial for a repeating chain knot $K$ with
  $3n$ crossings  as shown in Fig. 3(a).
This repeating  knot is a special 
 case  of a general repeating tangles  whose Jones polynomial
has been evaluated by Kauffman and Wu \cite{kw}.
Here, 
using
the Potts model formulation described in Sec. 2,
we give an independent derivation of the Jones polynomial for the knot.
Note that the  knot is the trefoil when $n=1$.
 
\medskip
First we shade alternate faces of
$K$ as  in Fig. 3(b) to yield an unoriented knot diagram with
crossing signs  shown.  Clearly, we have
\bea
&& n_+=2n-2,\hskip .5cm  n_-=n+2 \nonumber \\
&& m_+=0, \hskip 1.3cm m_-=3n, 
 \label{mn}
\eea
and  the Potts spins interact with a uniform interaction  
 $K_-$.  The resulting Potts model lattice is shown in Fig. 3(c) which has $M=2n+1$ sites.

\begin{figure}[htbp]
\center{
\epsfig{figure=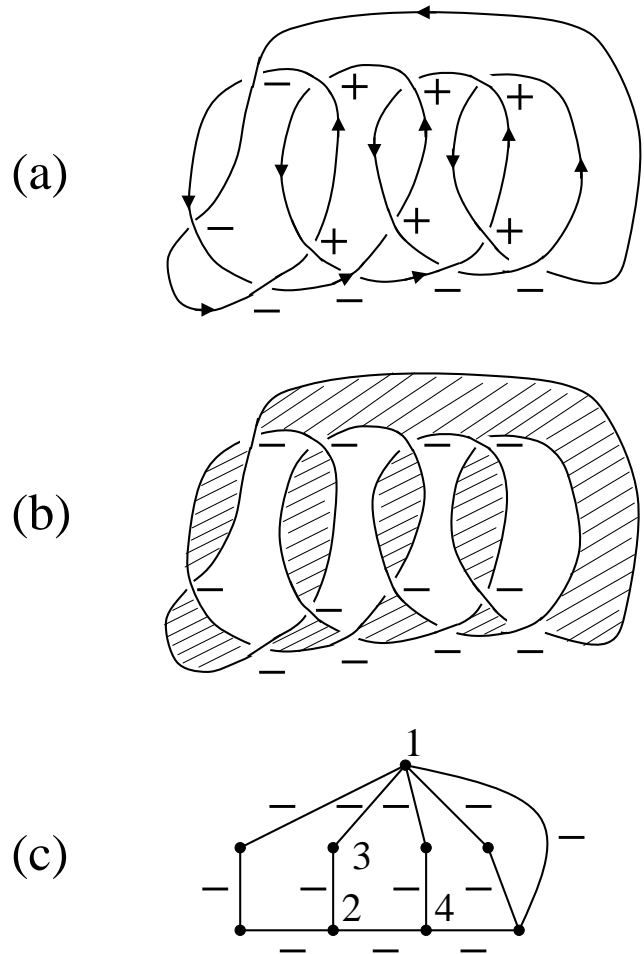,width=4.4in}}
\vskip -.4in
\caption{(a) A repeating chain knot with $3n$ crossings for $n=4$. Signs of
      crossings are those associated with oriented knots.
      (b) A shading of alternate faces.  Signs of crossings are those determined
         by the topology of the shading relative to the crossings (see text).
      (c) The resulting Potts model lattice for the face shading in (b).
   Labellings 1, 2, 3, 4 are those given in (\ref{rec}).}
\label{fig3}
\end{figure}

\medskip
To evaluate the partition function $Z_n(q;e^{K_-})$ for this
Potts model, we 
define 
\be
F_{m}(\sigma_1, \sigma_2) = a_{m} + b_{m} \delta_{\rm Kr}(\sigma_1, \sigma_2),
\hskip 1cm m=0,1,2,...
\ee
and the recursion relation
\bea
 F_{m+1}(\sigma_1, \sigma_2) &=& \Big[ \sum_{\sigma_3=1}^q  
B(\sigma_1, \sigma_3)B(\sigma_3, \sigma_2)\Big]  \cdot
\Big[ \sum_{\sigma_4=1}^q  B(\sigma_2, \sigma_4) F_{m}(\sigma_4, \sigma_1)\Big] \nonumber \\
&& \hskip 3cm 
\hskip 1cm m=0,1,2,... \label{rec}
\eea
obtained from a perusal of Fig. 3(c),
with the  initial condition $(a_0, b_0)=(0,1)$.  
The Potts partition function is then given by
\be
Z(q;e^{K_-}) = \sum_{\sigma_1=1}^q\sum_{\sigma_2=1}^q  F_{n}(\sigma_1, \sigma_2)
     = q (q a_n +b_n) \label{pottspf}
\ee
It therefore remains to compute $a_n$ and $b_n$.

\medskip
To compute $a_n$ and $b_n$, we rewrite
the recursion relation (\ref{rec}) as
\be
\pmatrix{a_{m+1} \cr b_{m+1}} = \pmatrix{ T_{11} & T_{12} \cr T_{21} & T_{22}\cr}  
\pmatrix{a_{m} \cr b_{m}}, \hskip 1cm m=0,1,2,... \label{transferm}
\ee
where
\bea
T_{11}&=& (q-2+2e^{K_-}) (q-1+e^{K_-}) \nonumber \\
T_{12}&=& q-2+2e^{K_-})  \nonumber \\
T_{21}&=& (e^{K_-}-1)^2 (q-1+e^{K_-}) \nonumber \\
T_{12}&=& e^{K_-}(1-e^{K_-})^2 -(1-e^{K_-})(q-2+2e^{K_-}).
\eea
Let $\lambda_1$ and $\lambda_2$ be the eigenvalues of the $2\times 2$ matrix in (\ref{transferm}).
After some straightforward manipulation, we arrive at the expression
\be
\pmatrix{a_{n} \cr b_{n}\cr} = {1\over {\lambda_1-\lambda_2}}
\pmatrix{a_0 (\lambda_1^{n+1} -\lambda_2^{n+1}) + (b_0T_{12} -a_0T_{22})
  (\lambda_1^{n} -\lambda_2^{n}) \cr
b_0 (\lambda_1^{n+1} -\lambda_2^{n+1}) + (a_0T_{21} -b_0T_{11})
  (\lambda_1^{n} -\lambda_2^{n}) }
\ee
 
\medskip
Now we substitute with  $(a_0, b_0)=(0,1), q=t+2+t^{-1}$ and $e^{K_-} = -t$,
and obtain the eigenvalues
\bea
\lambda_1 &=& -t^{-1}(1+t)^2 \nonumber \\
\lambda_2 &=& t^{-2}(1+t)^2 (1-t+t^2-t^3).
\eea
This leads to the partition function
\be
Z(q;e^{K_-}) = {{(1+t^{-1})^{2n}}\over {1+t^2-t^3}}
 \Big[ (-t)^{n+2} +(1-t^3)(1-t+t^2-t^3)^n  \Big] \label{zt}
\ee
Substituting (\ref{zt}) into (\ref{V}) with  $M=2n+1, N=3n$ and 
making use of (\ref{mn}),
we obtain
  the Jones polynomial 
\be
V_K(t) = {{t^{-(n+3)}}\over {1+t^2-t^3}}
  \Big[ t^{n+2} -(t^3-1)(t-1)^n(t^2+1)^n  \Big], \hskip .5cm n=1,2,...  \label{repk}
\ee
 The zeroes of $V_K(t)$  are now the roots of the equation
\be
t^{n+2} -(t^3-1)(t-1)^n(t^2+1)^n = 0 \label{roots}
\ee
minus the three roots of $1+t^2-t^3=0$.

\medskip
We have evaluated the roots of (\ref{roots})
in the complex $t$ plane numerically  for $n\leq 23$, and present the results 
  for 
typical values of $n$ in Fig. 4, and 
 for all $n\leq 23$ in Fig. 5(a). 
It is  seen that  the zeroes
  generally fall on three closed curves centered about the  points
$1$, $i$ and $-i$.  In addition, there are  two isolated zeroes
located near $t_+$ and $t_-$ and approaching $t_\pm$ as $n\to \infty$.
 
\begin{figure}[htbp]
\center{
\epsfig{figure=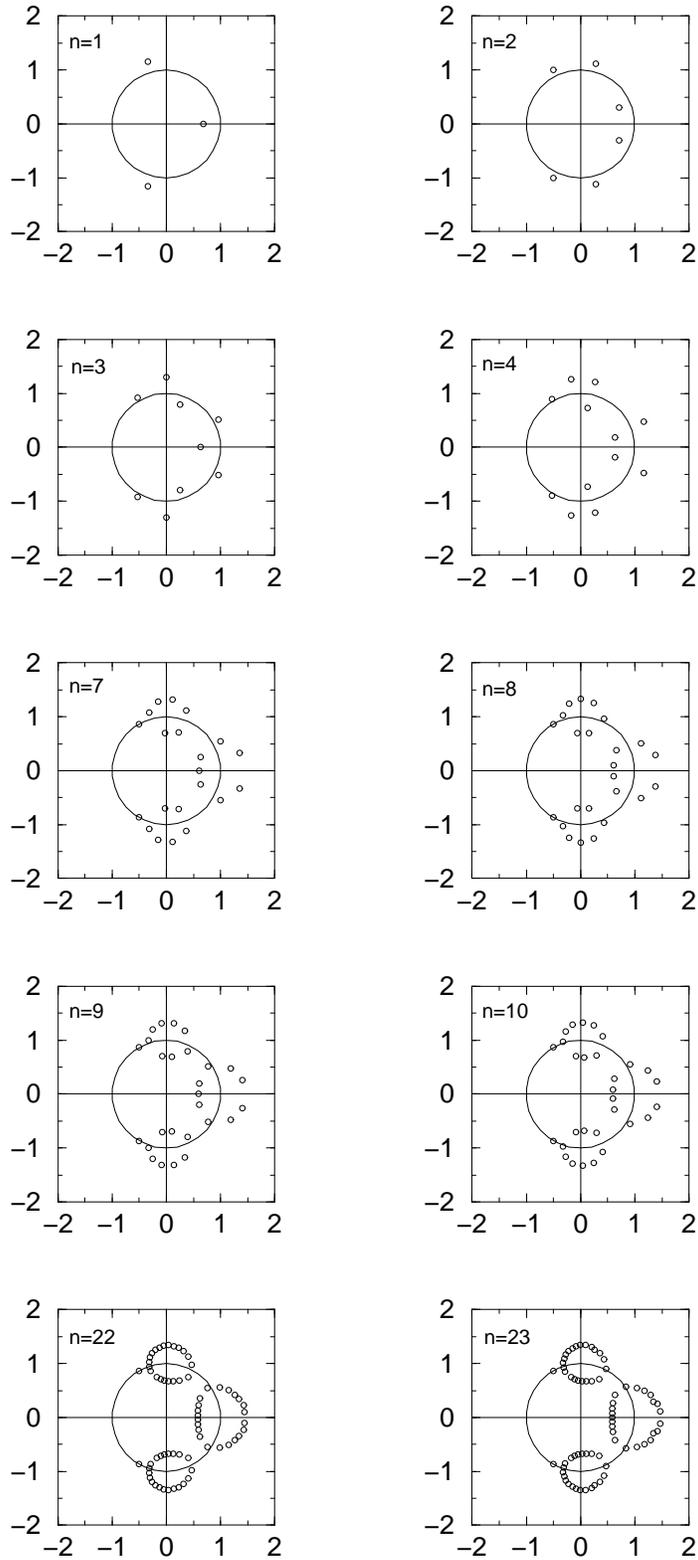,width=3.6in}}
\vskip -1.7in
\caption{Distribution of zeroes for the repeating chain knot for fixed $n$.}
\label{fig4}
\end{figure}

To determine the loci of the closed curves, we   
solve $t^n$ from  (\ref{roots}) and obtain, after taking the $n$-th root,
\be
{{(t-1)(t^2+1)}\over t} = \Bigg| {{t^2} \over {t^3-1}} \Bigg|^{1/n} \o^\alpha, 
\hskip 1cm \alpha = 1,2,...,n. \label{rep1}
\ee
Thus, in the limit of $n\to\infty$ and $t^3 \not= 1$, the zeroes
are distributed continuously on the three loci which are  the  roots of the cubic equation  
\be
(t-1)(t^2+1) =  t\  e^{i\theta}, \hskip 1cm 0\leq \theta < 2 \pi.  \label{loci}
\ee
 The loci (\ref{loci}) coincides with the crossover of the two eigenvalues, namely,
\be
\big| \lambda_1 \big| = \big| \lambda_2 \big|
\ee
so that the Jones polynomial is nonanalytic at (\ref{loci}). The three loci are 
 plotted  in Fig. 5(b).

\begin{figure}[htbp]
\center{
\epsfig{figure=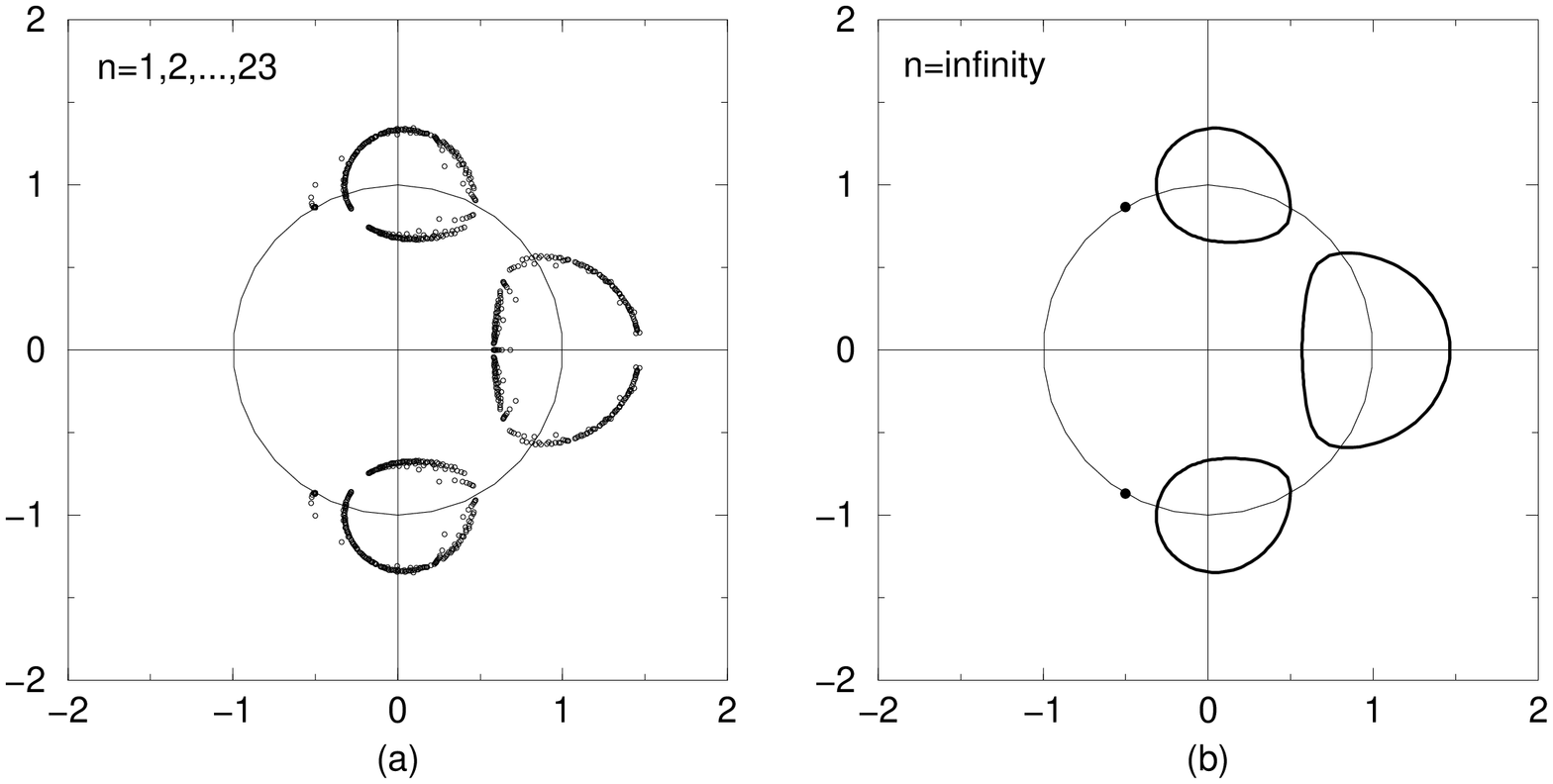,height=4.6in,angle=-90}}
\vskip -.7in
\caption{(a) Distribution of zeroes for the repeating chain knot for $n\leq 23$.
         (b) Distribution of zeroes for the repeating chain knot of in the limit of
        $n\to\infty$. The three loci are given by (\ref{loci}).}
\label{fig5}
\end{figure}

In the neighborhood of   $t_\pm$, we have $t^3\sim 1$ and
write 
$t=t_\pm +\Delta$ for some small $|\Delta|$.   
Then, to the leading order of $|\Delta|$  (\ref{rep1}) reads
\be
\pm \sqrt 3 i = |3\Delta|^{-1/n}\ t \ \o_n^\alpha. \label{isol}
\ee
Taking the absolute value of both sides, 
  (\ref{isol})  has  a solution  
\be
|\Delta|\sim 3^{-(n+2)/2}
\ee
consistent with the assumption of   $|\Delta|$ being small.
This leads to 
an additional isolated root near $t_+$ (and $t_-$) at a distance of the order of 
$3^{-(n+2)/2}$,
a fact  borne out by our numerical results.  The
two isolated roots   approach the  points $t_\pm$
on the unit circle as $n$ increases, and merges onto the unit circle in the limit of $n\to\infty$.
The average distance of the roots on the loci to the unit circle is 
 $0.2651$.
  
\section{Summary and discussions}
We have investigated the distribution of zeroes of the Jones polynomial
for all prime knots with $N\leq 10$ crossings, the torus knot, and a
repeating chain knot of $n$ sections for general $n$.
It is intriguing  to note that the zero distribution for  prime knots
shown in  Fig. 1
 resembles that of a Julia set. 
It would be of interest to further clarify the possibility of this connection.
 
We have also given the explicit relation which connects the Jones polynomial
to an associated Potts model, and used it to evaluate the Jones polynomial
of the repeating chain knot.
While the examples considered here are limited in scope,
it is hoped that further studies along this line can lead to further
 revelation of the
analytic properties of the Jones polynomial.

 \medskip
{\it Note added}: 
After the completion of this work we have learned that 
X.-S. Lin has studied
zeroes of several Jones polynomials \cite{lin}. His findings on the
torus knot are in agreement with ours.  We have also been informed by R. Shrock that
  zeroes of the Jones polynomial for 
certain repeating chain of alternate knots have been studied via its
 connection with the Tutte polynomial \cite{sc}.

\section*{Acknowledgments}
We would like to thank L. H. Kauffman for suggesting the
investigation of 
 the repeating chain knot
considered in Sec. 5.  We also thank 
B. Hu for comments on the possible connection
with Julia sets, and  H. Y. Huang and
W. T. Lu for technical assistance.
   This research is supported in
part by National Science Foundation Grant DMR-9980440.

\end{document}